\documentclass[twocolumn,showpacs,amsmath,amssymb]{revtex4}
\usepackage{graphicx}
\usepackage{epsfig,amsmath,natbib}
\begin{document}
\title[]{Cyclostationary shot noise in mesoscopic measurements}%
\author{Leif Roschier}
\author{Tero T. Heikkil\"{a}}
%\email{Tero T. Heikkila@hut.fi}
\author{Pertti Hakonen}
\affiliation{Helsinki University of Technology, Low Temperature
Laboratory, P.O.BOX 2200, FIN-02015 HUT, Finland} 
\begin{abstract}
{We discuss theoretically a setup where a time-dependent current
consisting of a DC bias and two sinusoidal harmonics is driven
through a sample. If the sample exhibits current-dependent shot
noise, the down-converted noise power spectrum varies depending on
the local-oscillator phase of the mixer. The theory of this
phase-dependent noise is applied to discuss the measurement of the
radio-frequency single-electron transistor. We also show that this
effect can be used to measure the shot noise accurately even in
nonlinear high-impedance samples.}
\end{abstract}

\maketitle

The dominating noise mechanism in mesoscopic samples at low
temperatures is shot noise. In some cases, it is the limiting
factor for the measurement sensitivity, but it may be the measured
quantity itself as it, contrary to the thermal noise, contains
information about the sample complement to that of the average
current \cite{blanter00}. Many of the interesting predictions for
noise have been obtained for nonlinear elements (with
voltage-dependent response) whose resistance is typically in the
range of k$\Omega$ or more. However, measurement of shot noise in
such samples is not always straightforward as the excess noise
added by the amplifiers depends on the sample impedance, and thus
on the applied voltage. An important property of shot noise is
that it is typically proportional to the average current. In this
paper we exploit this property and show that with adiabatic
cyclostationary driving, generalizing the treatment in
Refs.~\cite{niebauer91,rakhmanov01}, one may modulate the noise by
using a few first harmonics of the base frequency. We show that
this can be used to improve the sensitivity of the radio-frequency
single electron transistor (RF-SET) by some fraction. We also show
that the use of the phase dependence of the noise allows the noise
to be measured without a precise knowledge of the amplifier noise.

\begin{figure}[ht]
\begin{center}
\includegraphics[width=0.7\columnwidth]{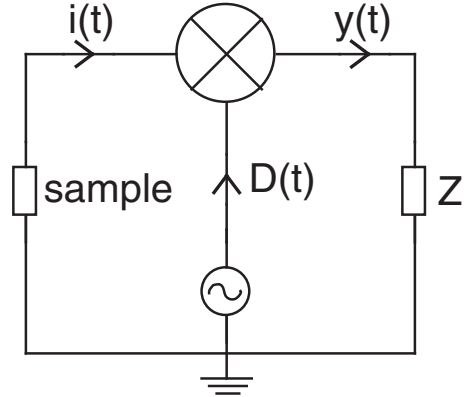}
\label{skema} \caption{The general setup. Current $i(t)$ flows
through the sample. Output current is $y(t)=i(t)\times D(t)$,
where $D(t)=D_0 \sin (\omega_0 t+\phi)$. It is converted into a
signal voltage using detection circuitry. }
\end{center}
\end{figure}

Throughout the paper we consider a situation illustrated in
Fig.~\ref{skema}. We limit to the case where current $i(t)=I_0+I_1
\sin (\omega_0 t) + I_2 \cos (2\omega_0 t)+n(t)$ flows through the
sample. We assume that $\langle i(t) \rangle= I_0+I_1 \sin
(\omega_0 t) + I_2 \cos (2\omega_0 t) \equiv I(t)$ and noise
$\langle n(t) \rangle =0$. For a low enough driving frequency
$\omega_0$, shot noise adiabatically follows the absolute value of
the current. Moreover, in this low-frequency limit, the noise is
white. This means that
\begin{equation}
2 \int_{-T/2}^{T/2} d\Delta t e^{-i\omega \Delta t} \langle
n(t+\Delta t/2) n(t-\Delta t/2)\rangle = 2eF|I(t)|, \label{tahti}
\end{equation}
independent of the frequency $\omega$. To account for a
general shot noise source, we introduced the Fano factor $F$.
Finite time Fourier transform of Eq.~(\ref{tahti}) with
$e^{-i(\omega-\omega')t}$ yields
\begin{equation}
\langle \tilde{n}(\omega)\tilde{n}^*(\omega')\rangle
=eF\tilde{I}_A(\omega-\omega').\label{nn}
\end{equation}
The terms $\tilde{n}(\omega)$ and $\tilde{I}_A(\omega)$  are
defined using the finite time Fourier transforms:
$\tilde{g}(\omega )\equiv \int_{-T/2}^{T/2} e^{-i\omega t} g(t)
dt$ and $\tilde{g}_A(\omega )\equiv \int_{-T/2}^{T/2} e^{-i\omega
t} |g(t)| dt$. It is assumed that $T=2 N\pi/\omega_0$, where $N
\gg 1$ is an integer. Equation (\ref{nn}) is derived for a diode
($F$ = 1) in the case $I(t)>0$ in Refs. ~\cite{niebauer91,
rakhmanov01} and the generalization to the case with arbitrary
sign of $I(t)$ is straightforward.

The goal is to find the power spectral density $S_y(\omega )
\equiv \lim_{T\rightarrow \infty} 2\frac{\langle |y(\omega )|^2
\rangle }{T}$ for $y(t)=D(t) \times i(t)$, where $D(t)=D_0 \sin
(\omega_0 t+\phi )$. The factor two accounts for the contributions
of both negative and positive frequencies. By noting that
\begin{widetext}
\begin{equation}
\tilde{y}(\omega)=\frac{D_0}{2i} \left[ e^{i\phi }\Big(
\tilde{I}(\omega - \omega_0)+\tilde{n} (\omega - \omega_0)\Big) -
e^{-i\phi }\Big( \tilde{I}(\omega + \omega_0) + \tilde{n} (\omega
+ \omega_0)\Big) \right]
\end{equation}
and by using Eq. (\ref{nn}), we find in the band $0 \leq \omega <
\omega_0$
\begin{equation}
\langle \tilde{y}(\omega) \tilde{y}^*(\omega) \rangle =
\frac{D_0^2}{4} \Big\{ \frac{I_1^2}{2} \Big( 1 +\cos (2 \phi )
\Big) |\delta_T(\omega)|^2 +F\Big( 2 \tilde{I}_A(0)-e^{2i\phi }
\tilde{I}_A(-2\omega_0 ) - e^{-2i\phi } \tilde{I}_A(2\omega_0
)\Big)\Big\}, \label{yy}
\end{equation}
\end{widetext}
where $\delta_T (\omega) =1/2\pi \int_{-T/2}^{T/2} e^{-i\omega t}
dt$. The first term in (\ref{yy}) is usually taken as the
downconverted carrier for the signal and the other terms as noise.
It is to be noted that only the DC and $2\omega_0$ components of
$\tilde{I}_A(\omega )$ contribute to the noise. It is thus enough
for the noise calculations to find the Fourier series expansion
$|I(t)| \simeq I_{0A} + I_{1A} \sin (\omega_0 t+\theta_A )+I_{2A}
\cos (2 \omega_0 t)$, where $I_{1A}$ and $\theta_A$ have no effect
on the noise. By using (\ref{yy}) we find the power spectral
density of the noise in the band $0< \omega <\omega_0$
\begin{equation}
S_y(\omega) = e F D_0^2\Big[I_{0A}-\frac{I_{2A}}{2}\cos{2\phi}
\Big]. \label{sy}
\end{equation}
Equation (\ref{sy}) is the same as in Refs.~\cite{niebauer91,
rakhmanov01} except that we have defined it in terms of Fourier
expansion coefficients $I_{0A}$ and $I_{2A}$ in order to handle
alternating currents.

\hyphenation{electro-meter} We first apply Eq.(\ref{sy}) to the
analysis of the RF-SET, that is the most sensitive electrometer
known today~\cite{schoelkopf,aassime}. In practice, its
sensitivity is limited by the noise of the following amplifier
\cite{roschier04}. In the near future when this noise will
probably be reduced from a few Kelvin to $\sim$100 mK ~\cite{muck}
the charge sensitivity will be set by the shot noise of the
tunnelling electrons through the single-electron transistor
(SET)~\cite{korotkov}. A metallic RF-SET is operated in two
different modes: In the superconducting case the best sensitivity
is achieved with DC bias. In the normal (non-superconducting) case
the best sensitivity is obtained without the DC bias. In this
latter case, the current $I(t)$ through the SET is given by $I_1
\sin (\omega_0 t)$ and we find $|I(t)| \simeq I_{0A}+I_{2A}
\cos(2\omega t)$, where $I_{0A}=I_12/\pi$ and
$I_{2A}=-4I_1/(3\pi)$.
The operation of the RF-SET is based on amplitude modulation,
because the signal is due to the resistance variation in time.
This means that the largest signal component comes out when
$\phi=0$ and the shot noise $S_y$ is at maximum according to Eq.
(\ref{sy}). In contrast, if the RF-SET could be operated in a
phase variation mode such that the signal component would be in
the $\cos (\omega_0 t)$ quadrature ($\phi=\pi /2$), then the shot
noise contribution would be by a factor of two lower in power.

In the other measurement mode, the SET is DC biased and the
current through it may be approximated by $I(t)=|I(t)|=I_0+I_1
\sin (\omega_0 t)=I_{0A}+I_{1A} \sin (\omega_0 t)$. From Eq.
(\ref{sy}) it follows that the noise is set by the $I_0$ component
alone. Adding a component $I_2 \cos(2\omega_0 t)$ to the current,
the measured noise power spectrum may be lowered by some fraction.
If the signal component is negligibly affected by the $2\omega_0$-
component, the signal-to-noise ratio would also be enhanced.

\begin{figure}
\begin{center}
\includegraphics[width=0.8\columnwidth]{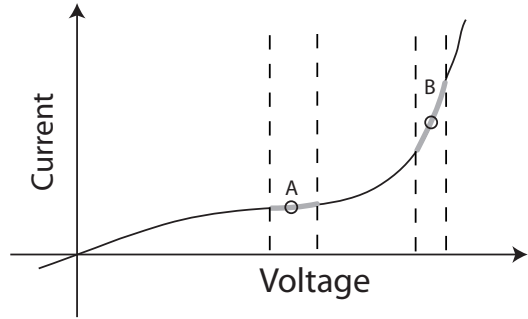}
\label{generaliv} \caption{An arbitrary nonlinear IV curve
illustrating the problem of the noise measurements. Regions A and
B have different impedances and the amplifier noise contribution
to the total noise is different at these points. In $2\omega_0$
modulation, the sample is DC biased (marked by circles) and a
signal $I_{2}\cos (2 \omega_0 t)$ is added (marked as gray area).
By measuring the noise in both quadratures by varying mixer phase
$\phi$, one may extract the shot noise contribution to the total
noise.}
\end{center}
\end{figure}

In some cases, noise reveals additional information that cannot be
extracted from IV curve measurements.
For example, in the subgap regime of
normal-insulator-superconductor contact, shot noise could be used
as a tool to determine whether conduction is due to quasiparticles
(thermal excitation or leakage current) or due to Andreev
reflection. However, noise measurements at low frequencies of such
nonlinear high-resistance samples with conventional techniques are
possible only with special care~\cite{cron01} provided that
1/$f$-noise is negligible. Alternatively, measurements have been
carried out on very large samples in order to circumvent the
problems~\cite{lefloch03}.
% -----
A characteristic illustration is depicted in Fig.~\ref{generaliv}:
IV curve is measured, but noise contribution of the sample cannot
be extracted from the total measured noise. This is due to the
dominating amplifier noise variation, that is difficult to know
accurately, with respect to the sample impedance. In general,
measurement of the shot noise in mesoscopic samples with high
impedance levels ($>$ 1 k$\Omega$) is difficult due to the small
signal levels.

In the following we propose a technique to use $2\omega_0$
modulation in order to measure the shot noise. By $2\omega_0$
modulation technique we mean a setup, where local oscillator
signal in the mixer is of the form $D(t)=D_0 \sin (\omega_0
t+\phi)$ and there is a term $I_{2A}\cos (2\omega_0 t)$ in the
absolute value $|I(t)|$ of the current through the sample. In
order to avoid the 1/$f$ noise, we discuss noise measurements at
high frequencies, that is $f \sim 500$ MHz. In this frequency
range the signal is carried in coaxial cables. The restricted
cable size limits the characteristic impedance to values around
$Z_0 \sim$ 50 $\Omega$.

To be specific, we approach the problem by considering an example.
If the shot noise is to be measured from a current of 100 pA
assuming a Fano-factor $F$ of unity, the resulting noise
corresponds to a delivered power of $T_S\sim 2 eIZ_0$ to the
preamplifier. This equals 100 $\mu K$ in noise temperature. By
using the Dicke radiometer formula ~\cite{dicke46} for the
fractional error of the noise measurement $\Delta T/(T_N+T_S)
\sim1/\sqrt{B \tau }$, where $\tau$ is the measurement time and
$B$ the measurement bandwidth, it follows that with $\tau \sim 1$
s and $B\sim 10^8$ Hz the signal power of 100 $\mu$K may be
measured if the amplifier noise temperature $T_N$ is of the order
of 1 K. The practical problem is that $T_N$ depends in general
considerably on the impedance of the measured source
\cite{rothe,roschier04}. In order to extract the shot noise
contribution from the total measured noise, it would be necessary
to know the four noise parameters of the amplifier
~\cite{rothe,engberg} and their effect to the measured total noise
with an accuracy of 100 ppm. This is almost impossible in
practice. One solution is to use a series of cryogenic isolators
that have typically an isolation of 20 dB each. There is, however,
the problem that below 1 GHz isolators are bulky to fit into the
limited space of mK-cryostats.

The $2 \omega_0$ modulation technique can be used to partially
circumvent the problem considered above. As is obvious from
Eq.~(\ref{sy}), a change of the mixer local oscillator phase
$\phi$ varies the measured shot noise power by an amount $\sim
FeI_{2A}Z_0$ from the maximum to the minimum. This is an analogue
of the lock-in scheme frequently applied for the measurement of
the differential conductance. The point is that the
(phase-insensitive) amplifier noise does not directly depend on
$\phi$, but only through the possible bias dependence of the
sample impedance (which would make the amplifier noise
contribution partially follow the driving). The latter effect can
be reduced in the typically interesting limit where the sample
impedance exceeds the amplifier impedance, through the use of a
matching circuit whose characteristic impedance equals that of the
amplifier. In this case, provided the sample impedance does not
depend on the bias more strongly than the noise, the modulation of
the sample noise may be brought to dominate the environmental
contribution to the phase-sensitive noise. The advantage compared
to the direct measurement is that the contribution from the
environment in this case depends only on the nonlinearity of the
sample, whereas in the unmodulated measurement it is present even
for linear samples. In order to compare the results of the noise
measurement at different bias points, one needs to know the
impedance of the sample at these points. This may be found out by
measuring the reflection coefficient $\Gamma=\frac{Z-Z_0}{Z+Z_0}$
of the sample, where $Z$ is the sample impedance and $Z_0$ the
reflected wave impedance. In general, the total gain of the system
is not known accurately and it often varies on a time-scale of
hours. Therefore, the measurement produces a differential result
$dS_y/dI_{2A}=G F(I_{0A})$ of the possibly bias-dependent Fano
factor $F(I_{0A})$ with one unknown multiplicative gain factor
$G$. If there is, however, one measured point that has a well
established result, for example $F=1$ as is the case in typical
tunnelling structures at high voltages, the measurement result at
this point can be used as a calibration to extract gain $G$, and
the Fano factor can be found out over the whole measurement range.

As a conclusion, we have discussed the theory of a doubly
modulated shot noise. We applied the outcome, Eq.~(\ref{sy}), to
the zero DC bias RF-SET and concluded that there is by a factor
two more noise in the in-phase quadrature than in the out-of-phase
quadrature. We suggested the use of the $2\omega_0$ modulation in
order to reduce shot noise by some fraction in the case of the DC
biased RF-SET. We also argued that the $2\omega_0$ modulation may
be used to extract the shot noise contribution from the total
noise. This may be applied, e.g., to determine Fano factors from
high-impedance samples at current levels below 1 nA using an
amplifier with noise temperature $T_N\sim$ 1 K, a measurement
bandwidth of 10$^8$ Hz, and a measurement time of 1 s.

\end{document}